\documentstyle[11pt,paspconf,epsf]{article}

\begin{document}

\title{Far infrared and Far ultraviolet emissions of galaxies: luminosity 
functions and selection effects. Implications for the future NGST and ALMA 
observations}

\author{V. Buat and D. Burgarella }
\affil{IGRAP, Laboratoire d'Astronomie Spatiale, Marseille FRANCE }



\begin{abstract}
The FIR(60 $\mu$m) and UV (0.2 $\mu$m) emissions of 
individual star forming galaxies  are compared to the mean 
properties of the local Universe. Almost all the galaxies 
exhibit a FIR to UV flux ratio larger than the ratio of 
the FIR and UV luminosity densities. It is likely to be due 
to the contribution of low luminosity galaxies to the UV luminosity 
density: these galaxies are deficient in any survey. 

As the galaxies become brighter in FIR their FIR to UV flux ratio 
increases: (around 0.2 $\mu$m) 
it implies an increase of 0.5 mag  of extinction (around 0.2 $\mu$m) per decade 
of FIR (60 $\mu$m) luminosity.
This trend is extended and amplified in Ultra Luminous 
Infrared Galaxies observed at low and high redshift. 
The Lyman Break galaxies detected by their U dropout in 
the HDF seem to have less extreme properties more compatible
with the trend found in the nearby star forming galaxies.

The detection limits of 
the NGST and of the future large millimeter array (ALMA) in terms of FIR to UV 
ratios are presented.

\end{abstract}

\section{Introduction}
\vspace{-3mm}
 At high redshift the emission observed in the visible corresponds to the
ultraviolet range in the rest frame and the effects of the dust extinction can
be dramatic.  We need to know how to correct individual galaxies for extinction
but also how the properties of these individual cases can be extrapolated to the
entire population of the galaxies.  The FIR to UV flux ratio of galaxies is now
well recognized as a powerful indicator of the extinction (e.g.  Buat \& Xu
1996, Meurer et al.  1999).  The basic idea of the present work is to compare
the FIR (60 $\mu$m) and UV(0.2$\mu$m) emission of a sample of nearby galaxies
for which selection biases are well known.  The aim is to study how much these
individual galaxies trace the mean properties of the local universe.  The second
step is to compare the properties of this sample to individual and more distant
galaxies so far observed in FIR and UV (rest frame).

\section{The FIR(IRAS)/UV(FOCA) ratio of nearby galaxies}
\vspace{-3mm}
\subsection{ The IRAS/FOCA sample of nearby galaxies}
 
We have cross-correlated Far Infrared (IRAS) and Far Ultraviolet (FOCA, Donas et
al.  1995) observations of galaxies to construct {\it a sample of FIR selected
galaxies with a UV (0.2 $\mu$m) observation}.

 We have considered the 22 fields ($\sim$ 70 square degrees) observed in UV by
the FOCA experiment and at present calibrated.  We started from the IR
detections for which much data are available and search for their UV
counterparts.  83 spiral and irregular galaxies have been observed and securely
measured both in UV and FIR.

\subsection{The FIR to UV ratio in the IRAS/FOCA sample: Comparison 
with the luminosity densities}

In Figure 1 is reported the ratio of fluxes at 60 $\mu$m and 0.2 $\mu$m, $\rm
F_{60}/F_{0.2}$ as a function of the luminosity of the galaxies at 60 $\mu$m.
($\rm F_\lambda = \lambda \cdot
f_\lambda $ where $\rm f_{\lambda}$ is a flux per unit wavelength).
There is a clear trend for a larger $\rm F_{60}/F_{0.2}$ ratio
for brighter galaxies at 60 $\mu$m.  Using a moving median  and a linear fit on
the IRAS/FOCA data gives:  $\rm \log(F_{60}/F_{0.2}) = 0.31(\pm 0.07)~ \log
L_{60}-2.35(\pm 0.18)$. This fit is plotted as a solid line in Figure 1.
  
The FIR to UV flux ratio is a reliable tracer of the dust extinction (Buat \&
Xu 1996, Meurer et al.  1999, Witt \& Gordon 1999).  Using the calibration of
Meurer et al.  (1999) we find an extinction at $\rm \sim 0.2 \mu m$ increasing
from 0.8 mag for $\rm L_{60}= 10^8 L\odot$ to 2.5 mag for $\rm L_{60}= 10^{11}
L\odot$.  The calibration of Buat \& Xu (1996) gives extinctions $\sim$0.4 mag
lower (Buat et al.  1999).

\subsubsection{Comparison with the local luminosity densities}

The 60 $\mu$m local luminosity function and density at z=0 have been calculated
by Saunders et al.  (1990).  The 0.2 $\mu$m luminosity function and density have
been derived by Treyer et al.  (1998) at a mean z=0.15.  From these studies we
estimate the ratio of the local luminosity densities $\rm \rho_{60} /\rho_{0.2}$
at z=0:  $\rm \rho_{60}/\rho_{0.2} = 0.9 \pm 0.4$.  This value is reported in
figure 1 as a dotted line.  The ratio appears lower than almost all the ratios
found for individual galaxies and {\it the study of individual galaxies of our
sample does not lead to a reliable estimate of the mean FIR to UV ratio of the
local Universe}

\begin{table}
\caption{Contribution of the galaxies to the 0.2 $\mu$m and 60 $\mu$m 
luminosity 
function and to 
the 0.2$\mu$m and 60 $\mu$m 
luminosity density in the local Universe per decade of luminosity.  The
luminosity function is truncated at $\rm L=10^7 L\odot$($\rm h=H_0/100$)
} \label{tbl-1}
\begin{center}\scriptsize
\begin{tabular}{cccc}
$\rm log(L_{UV})$ & galaxies  &contribution & galaxies\\
solar unit& from the lum.func & to the lum.dens.&  IRAS/FOCA sample\\

\tableline
7-8$\rm h^{-2}$& 77.3$\%$&17 $\%$ &10$\%$\\
8-9$\rm h^{-2}$& 19.3$\%$&36$\%$&35$\%$\\
9-10$\rm h^{-2}$& 3.2 $\%$&36$\%$&55$\%$\\
10-11$\rm h^{-2}$& 0.2 $\%$&11$\%$&0$\%$\\
\tableline
\end{tabular}

\begin{tabular}{cccc}
\tableline
$\rm log(L_{FIR})$ & galaxies  &contribution & galaxies\\
solar unit&  from the lum.func & to the lum.dens.&  IRAS/FOCA sample\\
 
\tableline
7-8$\rm h^{-2}$& 45.4 $\%$&2$\%$&3$\%$\\
8-9$\rm h^{-2}$& 35.6 $\%$&15$\%$&17$\%$\\
9-10$\rm h^{-2}$&16.2 $\%$&40$\%$&57$\%$\\
10-11$\rm h^{-2}$& 2.7 $\%$&34$\%$&23$\%$\\
11-12$\rm h^{-2}$&0.1 $\%$&8$\%$&0$\%$\\
\tableline
\end{tabular}
\end{center}

\end{table}
\subsubsection{Comparison with the local luminosity functions}

An explanation for the discrepancy between the $\rm F_{60}/F_{0.2}$ ratio of
individual galaxies and the ratio of the local luminosity densities is that it
is not the same galaxies which form the bulk of the UV emission on one hand and
the FIR emission on the other hand.  This seems confirmed by the very different
shapes of the luminosity functions (Figure 2).  From Table 1 it is clear that
our individual galaxies do not truly sample the luminosity functions.  The
bright end of both luminosity functions is not represented in the IRAS/FOCA
sample because of the scarcity of these objects and the small statistics.  The
steepness of the faint end slope of the UV luminosity function implies a large
number of faint galaxies which largely contribute to UV luminosity density:
these galaxies are deficient in any survey and in particular in our IRAS/FOCA
sample.  The FIR luminosity function is better sampled in the sense that the
deficiency of low luminosity galaxies has less implications than in UV.  This is
due to the flatness of the FIR luminosity function at low luminosities.

\section{ Comparison with the FIR bright galaxies and the U-dropout galaxies of 
the Hubble Deep Field}
\vspace{-3mm}
\subsubsection{ FIR bright Galaxies}

We have also reported in Figure 1 the data available for nearby (Trentham et al.
1999) and high redshift (Hughes et al.  1998) Ultra Luminous Infrared Galaxies
(ULIGs) as well as the galaxies detected by ISOCAM in a CFRS field (Flores et
al.  1998).  All these FIR bright objects exhibit a larger $\rm F_{60}/ F_{UV}$
ratio than suggested by the extrapolation of the linear regression (solid line)
found for the IRAS/FOCA sample.  The nearby and high z ULIGs have very large
extinctions (from 7 to 11 mag) in UV as measured from the FIR to UV flux ratio
(Buat et al.  1999).The ISOCAM/CFRS galaxies exhibit a more moderate UV
extinction (2-5.5 mag) with a mean at 3.3 mag.  As a comparison the extinction
of Messier 82 at 0.2 $\mu$m estimated in the same way is 5.4 mag (Buat \&
Burgarella 1998).

\subsubsection{U-dropout, Lyman break galaxies}

Meurer et al.  (1999) have studied the U-dropout galaxies detected in the HDF to
their local starburst templates in terms of dust extinction and UV luminosity
distribution.  In order to compare these galaxies to our IRAS/FOCA sample we
have estimated their $\rm F_{60}$/$\rm F_{UV}$ ratio and their luminosity at 60
$\mu$m.\\ The 60$\mu$m luminosity is estimated from the Star Formation Rates
(SFRs) of the galaxies:  from the figure 6 of Meurer et al.  we deduce a range
in SFR 2-300 $\rm M_\odot/yr$.  The total infrared luminosity $\rm L_{IR}$ is
deduced from the relation $ \rm SFR=1.71~10^{-10}~ L_{IR} (L\odot)$ (Kennicutt
1998).  Adopting $\rm L_{IR}/L_{60}\simeq 1.4$ (Buat \& Burgarella, 1998; Meurer
et al.  1999), we obtain $\rm L_{60}$ in the range $8.5~10^9-1.3~10^{12}
L\odot$.\\ The $\rm F_{60}/ F_{UV}$ ratio is estimated from the extinction
deduced by Meurer et al.  with the shape of the UV continuum:  the range for
$\rm A(0.16\mu m)$ is 0.1-3 mag which translates to a $\rm F_{60}$/$\rm F_{UV}$
ratio comprised between 1.4 and -0.6 in log units (Buat et al.  1999, their
figure 1).\\ 
The location of the U-dropout galaxies in Figure 1 is within the dashed squared
area.  {\it The Lyman Break galaxies detected by their U dropout in the HDF seem
to have less extreme properties than ULIGs more compatible with the trend found
in the nearby star forming galaxies.}

\section{ The future NGST and ALMA observations}
\vspace{-3mm}
The {\bf New Generation Space Telescope(NGST)} coupled with the {\bf Atacama
Large Millimeter Array (ALMA)} will allow to obtain the FIR and UV rest-frame
emissions of galaxies in the early Universe.  We have tentatively estimated the
limiting luminosities reachable by these instruments at both wavelengths.  The
results are gathered in Table 2.  We have reported these limits in figures 1 and
2 for the redshifts $\sim$5 and $\sim$20.

The UV luminosity function is well sampled by the NGST up to z=20 whereas only 
FIR bright galaxies will be detected by ALMA at z$>$5. In terms of FIR to UV 
ratios, at z=5 all the range of $\rm F_{60}$/$\rm F_{UV}$ sampled by the objects 
reported in figure 1 will be reachable, at z=20 galaxies similar to ULIGs will 
be only marginally detected in UV.

\begin{table}
\caption{Detection limits of NGST and ALMA. The calculations are made for the 
four ALMA bands at 230, 350, 650 and 850 GHz
which correspond to a rest frame emission at 60 $\mu$m for z=20, 13.3, 6.7 and
4.8.  The corresponding NGST wavelengths corresponding to 0.2 $\mu$m rest frame
are 4.2, 2.9, 1.5 and 1.2 $\mu$m.  We assume point source distributions, a
signal to noise ratio of 3 and 1 hour exposure.} \label{tbl-2}
\begin{center}\scriptsize
\begin{tabular}{ccc}
 redshift &$\rm L_{lim}(0.2\mu m) (solar units)$&$\rm L_{lim}(60\mu m)$ (solar 
units)\\

\tableline
4.8& $6.2~10^7~h^{-2}$ &$2.7~10^{10}~h^{-2}$\\
6.7& $10~10^7~h^{-2}$&$2.5~10^{10}~h^{-2}$\\
13.3&$23.8~10^7~h^{-2}$&$1.25~10^{10}~h^{-2}$\\
20&$69~10^7~h^{-2}$ &$1.33~10^{10}~h^{-2}$\\
\tableline
\end{tabular}
\end{center}
\end{table}

\begin{figure}
\plotfiddle{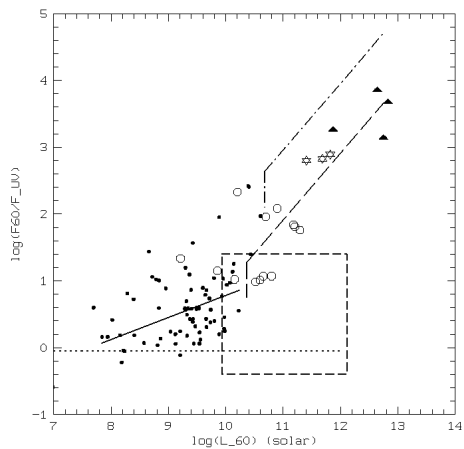}{5.5truecm}{0}{25}{25}{-100}{0}
\caption{ The ratio of the emission at 60 and 0.2 $\rm \mu m$ as a
function of the luminosity at 60$\rm \mu m$. The IRAS/FOCA sample is plotted as 
dots,  the linear fit is represented by the solid line. The ratio of the local 
luminosity
densities $\rho_{60}/\rho_{0.2}$ is reported as a dotted horizontal line. 
FIR bright and U-dropout 
galaxies are overplotted: CFRS/ISOCAM galaxies as empty circles, nearby ULIRGs 
as  stars and high z ULIRG as filled triangles. {\bf ALMA/NGST} detection 
limits are represented as a dash-dotted line (z=5) and a long dashed  line 
(z=20)} 
\end{figure}

\begin{figure}
\plotfiddle{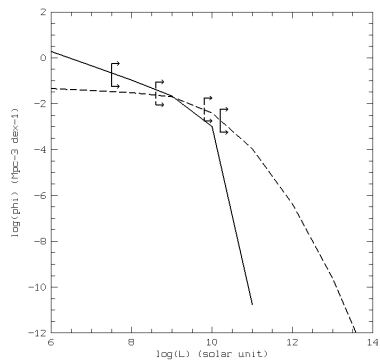}{4truecm}{0}{20}{20}{-100}{0}
\caption{ The luminosity functions at 0.2 $\mu$m (solid curve) and 60 
$\mu$m (dashed curve). The detection limits of the NGST and ALMA for $\rm z=5$ 
(solid arrows) and $\rm z=20$ (dotted arrows) are reported  on both curves.} 
\end{figure}

\end{document}